\documentclass[journal]{IEEEtran}

\usepackage[utf8]{inputenc}
\usepackage{comment}
\usepackage{amsmath,amsmath,bm,amsfonts}
\usepackage{graphicx}
\usepackage{textcomp}
\usepackage{booktabs}
\usepackage{cite}
\usepackage{pifont}

\newcommand{\xvi}[1]{\mathbf{x}_{#1}}
\newcommand{\zvi}[1]{\mathbf{z}_{#1}}
\newcommand{\cvi}[1]{\mathbf{c}_{#1}}

\newcommand{\zvSi}[1]{\underline{\mathbf{z}}_{#1}}

\newcommand{\Xm}[1]{\mathbf{X}_{#1}}
\newcommand{\Zm}[1]{\mathbf{Z}_{#1}}
\newcommand{\Am}[1]{\mathbf{A}_{#1}}

\newcommand{\cmark}{\ding{51}}%

\usepackage{color, colortbl}
\definecolor{red}{rgb}{0.9,0.3,0}
\definecolor{Gray}{gray}{0.95}

\title{Regularizing Contrastive Predictive Coding for Speech Applications }
\author{Saurabhchand Bhati$^{\dagger}$, Jes\'us Villalba$^{\dagger,\ddagger}$, Piotr \.Zelasko$^{\ast}$, Laureano Moro-Velazquez$^{\dagger}$, Najim Dehak$^{\dagger,\ddagger}$ \\
$^{\dagger}$Center for Language and Speech Processing, Johns Hopkins University, USA \\
 $^{\ddagger}$Human Language Technology Center of Excellence, Johns Hopkins University, USA  \\
 $^{\ast}$Meaning.Team Inc, USA \\
 \{sbhati1\thanks{This project was supported by NSF Award 1909075.},jvillalba,laureano,ndehak3\}@jhu.edu, pzelasko@meaning.team}

\begin{document}
\maketitle
\begin{abstract}
Self-supervised methods such as Contrastive predictive Coding (CPC) have greatly improved the quality of the unsupervised representations. These representations significantly reduce the amount of labeled data needed for downstream task performance, such as automatic speech recognition. CPC learns representations by learning to predict future frames given current frames. Based on the observation that the acoustic information, e.g., phones, changes slower than the feature extraction rate in CPC, we propose regularization techniques that impose slowness constraints on the features. Here we propose two regularization techniques: Self-expressing constraint and Left-or-Right regularization. We evaluate the proposed model on ABX and linear phone classification tasks, acoustic unit discovery, and automatic speech recognition. The regularized CPC trained on 100 hours of unlabeled data matches the performance of the baseline CPC trained on 360 hours of unlabeled data. We also show that our regularization techniques are complementary to data augmentation and can further boost the system's performance. In monolingual, cross-lingual, or multilingual settings, with/without data augmentation, regardless of the amount of data used for training, our regularized models outperformed the baseline CPC models on the ABX task. 

\end{abstract}
\noindent\textbf{Index Terms}: Self-supervised learning, zero resource speech processing, unsupervised learning, contrastive predictive coding 

\section{Introduction}
The speech signal contains information about linguistic units~\cite{hinton2012deep}, speaker
identity~\cite{dehak2010front}, the emotion of the speaker~\cite{kwon2003emotion}, etc. 
In a supervised scenario, the manual labels guide a strong classifier, such as a Deep Neural Network(DNN), to extract task-specific features. For example, paired with phone labels, a DNN learns to focus on extracting the acoustic information from speech and suppress the information about the speaker's identity. When paired with speaker labels, the DNN learns to focus on speaker information and suppress the phone information. 

In the unsupervised scenario, we do not have the guidance of manual transcriptions to select the relevant features and marginalize irrelevant information. A good speech representation becomes crucial for unsupervised systems' good performance~\cite{jansen2011efficient,badino2014auto,varadarajan2008unsupervised,huijbregts2011unsupervised,lee2012nonparametric,siu2014unsupervised,kamper2017segmental,bhati2017unsupervised,kamper2017embedded,bhati2018phoneme}. Learning good representations from unlabeled speech data could enable speech technologies in low-resource languages where limited or no amounts of labeled data are available~\cite{jansen2011efficient,badino2014auto,varadarajan2008unsupervised,huijbregts2011unsupervised,lee2012nonparametric,siu2014unsupervised,kamper2017segmental,bhati2017unsupervised,kamper2017embedded,bhati2018phoneme}. The goal of unsupervised representation learning is to capture the phone information and ignore the other sources of information, such as the speaker or channel. 

Self-supervised learning (SSL) methods have emerged as a promising technique for representation learning from unlabeled speech data~\cite{oord2018representation,schneider2019wav2vec,baevski2020wav2vec}. SSL has also been shown to be effective for learning representations in natural language processing~\cite{devlin2018bert,liu2019roberta} and computer vision~\cite{chen2020simple,caron2021emerging}. SSL methods learn representation by solving an auxiliary task for which the labels can be generated from the unlabeled data. For example, in Contrastive Predictive Coding (CPC)~\cite{oord2018representation}, a popular self-supervised method, the auxiliary task in next frame prediction. A CNN learns representations from the raw waveform, which are then fed into a recurrent neural network to generate contextual representations. 
The model is trained via noise contrastive estimation to correctly identify the correct next frame from a set of random frames given the contextual features.  

Self-supervised techniques such as CPC have drastically improved the quality of the representation learned from unlabeled data. CPC extracts a feature vector every 10 ms, i.e., 100 features/second, whereas underlying information, e.g., phones, change much more slowly.  
There have been several solutions to impose slow changes to the latent representations~\cite{chorowski2019unsupervised,bhati21_interspeech,chorowski2021aligned,cuervo2022contrastive,cuervo2022variable}. In this work; we propose regularizing constraints to impose slow changes in the latent representations. Ideally, the representation would stay constant within a phone and change abruptly at the phone boundaries. 

Self-expressing autoencoders (SEA)~\cite{bhati2020self} add an extra self-expressing constraint as a regularization term to the autoencoder.  
SEA tries to express the features extracted from the encoder as a linear combination of other features, thus enforcing the underlying information is shared among features. We modify the self-expressing constraint and use it to regularize the CPC training. We also propose Left-or-Right regularization (LorR) to constrain the nearby frames to be similar. LorR assumes a given frame shares a phone label with either the left or right frames. We add extra loss that minimizes the variance between the given frame and adjacent frames.

We pretrained the baseline CPC model and our proposed regularized CPC on Librispeech 100 hours and 360 hours portions. 
We evaluate the models on the ABX task of the Zerospeech 2017 benchmark, acoustic unit discovery, and the linear phone classification task on Librispeech 100 hours. We carry out a detailed hyper-parameter search to find the optimal weights for adding the regularization loss with the CPC loss. Experiments show that we outperform the baseline CPC on both the ABX and the linear phone classification tasks.

We also train CPC models on English, French, and Mandarin datasets from the Zerospeech 2017 challenge and evaluate the performance in monolingual, cross-lingual, and multilingual settings. We also evaluate the CPC models on ASR across ten languages from the common voice dataset. Across all these conditions, our regularization consistently improves the system's performance. 

Data augmentation has become an important part of both supervised~\cite{park2019specaugment} and self-supervised systems~\cite{chen2020simple,caron2021emerging,kharitonov2021data}. It allows us to train larger models by reducing overfitting. One major direction for SSL models is to learn augmentation invariant representation where a model is presented with two different augmented versions, and it must generate similar representations~\cite{chen2020simple,caron2021emerging}. 
Data augmentation combined with CPC significantly improves over baseline CPC~\cite{kharitonov2021data}. In this approach, the past and/or future audio signals are augmented with different augmentations, and the model is trained to predict the correct next frame. We show that our regularization techniques are complementary to the data augmentation techniques and can be used on top to improve the system performance further. 

The contributions of this work are summarized below:

\begin{itemize}
    \item We propose Left-or-Right regularization and Self-expression to enforce slowness constraints on the CPC features
    \item We show the proposed regularizations improve performance on ABX, linear phone classifier, acoustic unit discovery, and automatic speech recognition tasks
    \item We show our proposed regularization's work in monolingual, cross-lingual, and multilingual conditions 
    \item We show that the proposed regularizations are complementary and can be used with augmentation to improve the system's performance further
\end{itemize}

\section{Related Work}

In this work, we focus on unsupervised feature learning, where we do not have any labels for the training data. The ZeroSpeech challenges~\cite{versteegh2015zero,versteegh2016zero,dunbar2017zero,dunbar2019zero,dunbar2021zero} have been some of the significant drivers of progress in the unsupervised feature learning area. Same-different or ABX tasks have consistently been part of all the challenges that focus on evaluating the representations' quality. Before the popularity of SSL techniques, autoencoders~\cite{chorowski2019unsupervised,bhati2020self,kamper2019truly,renshaw2015comparison} were a dominant paradigm for learning representations. Autoencoders consist of two parts: an encoder which maps an input to a latent space, and a decoder which tries to reconstruct the input. The autoencoders are optimized to minimize the difference between the original and reconstructed inputs. Variational autoencoders (VAE)~\cite{kingma2013auto} proposed a different probabilistic interpretation of the feature learning framework. Vector Quantized VAE~\cite{van2017neural} replaced the continuous and stochastic latent vectors with deterministically quantized vectors. Since the quantization is not differentiable, a straight-through estimator is used to optimize the codebook used for quantization. 
Chorowski et al.~\cite{chorowski2019unsupervised} proposed a wavenet-autoencoder that encodes MFCC features into a latent space via a VAE and uses a wavenet decoder to reconstruct the original waveform.   

Contrastive Predictive Coding (CPC)~\cite{oord2018representation} and its variants~\cite{riviere2020unsupervised,van2020vector,bhati21_interspeech,chorowski2021aligned,cuervo2022contrastive,cuervo2022variable} have emerged as a popular choice for the representation learning task. Some of the best-performing solutions to previous challenges are CPC-based~\cite{van2020vector,chorowski2021aligned}. Even though some autoencoder-based methods tend to be more data efficient than CPC, given more data, CPC outperforms them on the ABX task~\cite{riviere2020unsupervised}. For the recent zero speech challenge, CPC has been the choice of feature extractor~\cite{dunbar2021zero}. However, quantizing the representations from the CPC degrades the performance on the ABX task~\cite{dunbar2021zero}. We chose the CPC as one of the baselines in the present work. 

There have been several attempts to impose slow changes on the unsupervised representations extracted from speech data.
Slow-feature analysis~\cite{wiskott2002slow} imposed a penalty on the rate of change of features to encourage slow changes in the features. A time-jitter regularization~\cite{chorowski2019unsupervised} was proposed to reduce the variability between adjacent embeddings of VQ-VAE. 
Chorowski et al. ~\cite{chorowski2019unsupervised2} added a penalty to divide the VQ-VAE features into a given number of piecewise-constant pieces. Although, this requires knowing the number of segments in advance. Kamper et al.~\cite{kamper2020towards} proposed a dynamic programming-based generalization of this approach to obtain phone segmentation from VQ-VAE and VQ-CPC features. Kamper et al.~\cite{kamper2022word} further extended this idea to apply dynamic programming (DP) iteratively to perform phone and word segmentation. The first step performs bottom-up phone discovery using DP and then performs symbolic word segmentation on top of the discovered units. The two stages are trained separately, i.e., word segmentation does not influence phone discovery. 

Bhati et al.~\cite{bhati21_interspeech} proposed Segmental CPC (SCPC): a hierarchical model which stacked two CPC modules operating at different time scales. The lower CPC operates at the frame level, and the higher CPC operates at the phone-like segment level. A simple differentiable boundary detector generates phone-like segments used for training the segment-level CPC. Both lower and higher levels CPCs are optimized jointly. They demonstrated that adding the second level CPC improves the phone boundary detection but degrades the phone class information present in the learned features~\cite{bhati2022unsupervised}. 
Chorowski et al.~\cite{chorowski2021aligned}  proposed Aligned CPC (ACPC), in which the model outputs a sequence of K $<$ M predictions that are aligned to the M upcoming representations. 
mACPC~\cite{cuervo2022contrastive} proposed a hierarchical model similar to SCPC where they used ACPC as the building blocks instead of CPC. mACPC obtained better feature discrimination than CPC, SCPC, and ACPC. mACPC further confirmed the tradeoff between boundary detection and classification performance. 

Hierarchical CPC (HCPC)~\cite{cuervo2022variable} stacked two CPC models and used reinforcement learning to generate the segment boundaries. They showed that it is possible to improve the classification performance of the features extracted from multilevel CPC by training the second-level CPC on segments extracted to optimize the next segment prediction task directly. There still seems to be some tradeoff between prediction quality and segmentation performance. HCPC obtains better phone discrimination than SCPC and mACPC but has lower phone segmentation performance than SCPC and mACPC. HCPC achieves state-of-art performance on the Zerospeech 2021 task.  

ACPC~\cite{chorowski2021aligned}, mACPC~\cite{cuervo2022contrastive} and HCPC~\cite{cuervo2022variable} all obtain better performance on ABX task than the baseline CPC. We compare our proposed regularization methods with all of them and show that we outperform them on the ABX task.  

\section{Regularized Contrastive Predictive Coding}
\subsection{Contrastive Predictive Coding}
\begin{figure}
    \centering
    \includegraphics[width=2.5in]{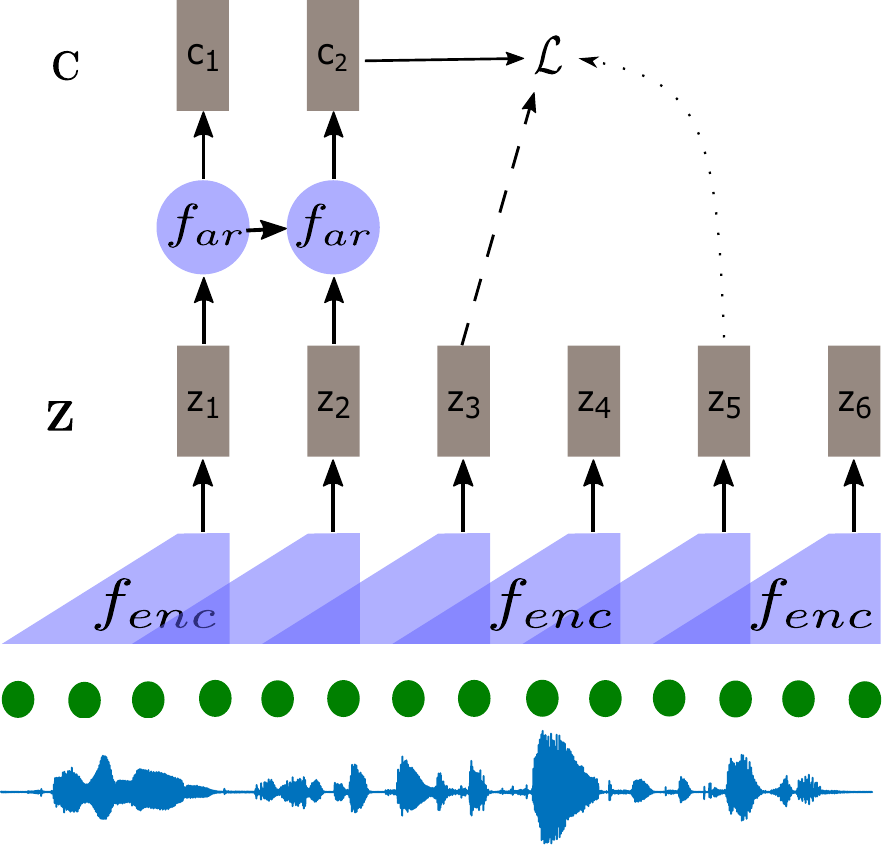}
    \caption{Overview of the CPC architecture. The solid line represents the reference frame; the dashed line shows the positive, and the dotted line shows the randomly sampled negative example.}
    \label{fig:cpc}
\end{figure}

Contrastive Predictive Coding (CPC)~\cite{oord2018representation} learns representations by predicting future feature frames from past frames. The architecture is shown in Figure~\ref{fig:cpc}. CPC can learn representations directly from raw speech waveforms. A convolution encoder, $f_{\mathrm{enc}}:\mathbf{X}\rightarrow\mathbf{Z}$, maps the audio waveform, $\mathbf{X}$ to latent spectral representations, $\mathbf{Z}(\in \mathbb{R}^{ d \times L}) = (\zvi{1},\zvi{2},...,\zvi{L})$. 
In the most common setting, each $d$-dimensional vector $\zvi{i}$ corresponds to a 30 ms audio frame extracted with a 10 ms shift. A recurrent neural network, $f_\mathrm{ar}: \mathbf{Z} \rightarrow \mathbf{C} $, extracts contextual representations $(\cvi{1},\cvi{2},...,\cvi{L})$ computed as $\cvi{i}= f_\mathrm{ar}(\zvi{i})$.
Given a reference context representation $\cvi{t}$ the model needs to identify the next frame  $\zvi{t+m}$ correctly from a set, $\mathcal{Z}_{t}$, of $K+1$  representations, which includes $\zvi{t+m}$ and $K$ distractors. $W_{m}$ is the linear transformation used for predicting $\zvi{t+m}$ from $\cvi{t}$. The overall loss is given as follows:

\begin{equation}
    \mathcal{L}_{\textrm{CPC}} = -\frac{1}{M} \sum_{m=1}^{M} \log \frac{\exp(\zvi{t+m}^{T}W_{m}\cvi{t})}{\sum_{\Tilde{\zvi{}} \in \mathcal{Z}_{t} } \exp(\Tilde{\zvi{}}^{T}W_{m}\cvi{t} )}    
\end{equation}

\subsection{Self-expressing Autoencoders}
In an autoencoder, an encoder, $\mathrm{enc}$ maps the input $\Xm{}$ into a latent space $\Zm{}(=\mathrm{enc}(\Xm{}))$ and the decoder $\mathrm{dec}$ takes the $\Zm{}$ and tries to reconstruct the input from it. The autoencoder is trained to minimize the reconstruction loss: $\Vert \Xm{} - \mathrm{dec}(\Zm{}) \Vert_{2}^{2}$.

The self-expressing autoencoders (SEA)~\cite{bhati2020self} introduce self-expressing constraints to encourage autoencoders to learn representations highlighting underlying phone information. 
The self-expressing constraint tries to make the embeddings of frames that belong to the same distribution to be as similar as possible and frames that belong to different distributions as dissimilar as possible.  

In SEA, an encoder maps the input features into an embedding space, and two decoders with shared parameters try to reconstruct the input as well as possible. One decoder tries to reconstruct the input from the embedding, and another tries to reconstruct from their self-expressed version. 

To compute the self-expressed version of the encoder outputs, we first compute the affinity matrix $\Am{}$, which captures the pair-wise cosine similarities between frames, $\Am{ij}$ denotes the cosine similarity between features $\zvi{i}$ and $\zvi{j}$. The last layer of the encoder is ReLU nonlinearly, and thus the $\zvi{}$ is non-negative, and therefore $\Am{}$ is non-negative. We then remove the contribution of the diagonal entries from $\Am{}$ and normalize each row, to sum up, to \textbf{1}.  
The self-expressed version, $\underline{\Zm{}}$, is given as 
\begin{equation}
\label{eq:z_sea}
\underline{\Zm{}} = \mathrm{rownorm}(\Am{}-\mathbf{I})\Zm{}
\end{equation}
where $\mathrm{rownorm}$ denotes the row normalization operation. The SEA is trained to minimize the reconstruction loss
\begin{equation}
    \mathcal{L}_{\textrm{SE}} = \Vert \Xm{} - \mathrm{dec}(\Zm{}) \Vert_{2}^{2} + \Vert \Xm{} - \mathrm{dec}(\underline{\Zm{}}) \Vert_{2}^{2}  
    \label{eq:loss}
\end{equation}

For the $i^{th}$ input, $\xvi{i}$, to minimize the reconstruction error, the decoder needs to reconstruct $\xvi{i}$ from both $\zvi{i}$ and $\zvSi{i}$.  $\zvSi{i}$ is a linear combination of features except for $\zvi{i}$. 
Other features should have the information present in $\zvi{i}$. 
Thus the information present in $\zvi{i}$ must be shared in other features.

\subsection{CPC with Self-expressive Constraint}
In the original SEA~\cite{bhati2020self}, to ensure the self-similarity matrix, $\Am{}$, is non-negative, the last layer in the encoder is ReLU non-linearity which restricts $\zvi{}$ to non-negative values. In the CPC architecture, the last layer of the encoder is ReLU as well. We calculate $\underline{\Zm{}}$ using~\eqref{eq:z_sea}. 

The CPC model does not use a decoder to learn representations. So, instead of minimizing
the difference between input and reconstructed input from the self-expressed embeddings,
we force the embeddings and their self-expressed versions to be as close as possible. The total loss of the model is given as
\begin{equation}
    \mathcal{L}_{\textrm{CPC+SE(ReLU)}} = \mathcal{L}_{\textrm{CPC}} + \lambda \Vert \Zm{} - \underline{\Zm{}} \Vert_{2}^{2} \;,
\end{equation}
where $\lambda$ is the regularization weight. 

\subsection{\textbf{L}eft-\textbf{or}-\textbf{R}ight (LorR) Regularization}
Assuming a phone consists of at least two feature frames, then any feature frame would have the same phone label as either the left or right frame. Most of the time, both left and right frames would have the same phone label. Only at the phone boundaries, which are much fewer than the total number of frames, the phone labels for left and right frames would be different. With the 30ms context size and 10 ms shift for the convolutional feature extractor, the minimum two frames phone assumption works out to be 40ms. Let's consider a sequence of four feature frames $\zvi{i-1}, \zvi{i}, \zvi{i+1}, \zvi{i+2}$. We want to constrain the features from the same phone to be as close as possible. However, in unsupervised scenarios, we do not know the phone labels. We minimize the minimum of variance between $\zvi{i-1}, \zvi{i}$ and $\zvi{i}, \zvi{i+1}$. At boundaries having a minimum allows the loss to be flexible, and it can choose the side with minimum variance, e.g., if there is a boundary at $\zvi{i}$, then the model could pick and minimize the variance between $\zvi{i}, \zvi{i+1} $ and vice-versa. In the middle of a phone, when both the left and right sides belong to the phone, the choice of side does not matter. We can extend this idea and try to enforce $w$ frames to be similar. The loss at $i_{th}$ feature is given as:

\begin{equation}
    \mathcal{L}_{i} = \min(\sum_{d} \mbox{Var}(\zvi{i-w+1:i}), \sum_{d} \mbox{Var}(\zvi{i+1:i+w}))
\end{equation}
Where $\sum_{d}$ denotes the sum of element-wise variance across $d$ dimensions. The total LorR loss is average across all the time indexes
\begin{equation}
    \mathcal{L}_{\textrm{LorR}} = \frac{1}{L} \sum_{i=0}^{L} \mathcal{L}_{i}
\end{equation}

The regularized CPC is optimized to minimize both the CPC loss and the LorR loss.

\begin{equation}
    \mathcal{L}_{\textrm{CPC+LorR}} = \mathcal{L}_{\textrm{CPC}} + \alpha \mathcal{L}_{\textrm{LorR}}
\end{equation}
where $\alpha$ is the regularization weight.

\section{Experiments}
\subsection{Tasks and Datasets}
We used LibriSpeech 100 hours, 360 hours, and the Zerospeech 2017 datasets for pretraining the CPC and regularized version of CPC. The Zerospeech 2017 train subset contains 45, 24, and 2.5 hours of data across English, French, and Mandarin, respectively. One common task for evaluating the quality of the representations is probing for the phone information by training a linear phone classifier. For the supervised training of the linear phone classifier, we used the train/test splits and force alignments for Librispeech-100h from~\cite{riviere2020unsupervised}. We also evaluated how well these representations can be clustered and mapped into discrete symbols.

Another common task is the ABX phone discrimination task.
ABX task measures the phone separability of the representations obtained from the feature extractor. Features from two instances of the same phone should be closer than two instances of different phones. For example, if phone instance $a$ and $x$ belong to the same phone class A and phone instance $b$ belongs to phone class B, then $d(a,x) < d(b,x)$ where $d$ is some distance metric. The ABX task was done in two modes: within speaker--when $a$, $b$, $x$ belong to the same speaker-- and across speaker--when $a$, $b$ belong to the same speaker, but $x$ belongs to a different speaker. For the ABX and linear phone classification task, we used the implementation provided by~\cite{riviere2020unsupervised}.

For the ABX task, we used the Zerospeech 2017 and Zerospeech 2021 challenge datasets for evaluations. Zerospeech 2017 test set contains the same speakers from the train set and an unknown number of new speakers appearing in 1-second, 10 seconds, and 120 seconds files. This allows us to measure the speaker invariance of the feature extraction system. More details about the dataset and the evaluation can be found in the challenge paper~\cite{dunbar2017zero}. The Zerospeech 2021 uses the standard Librispeech validation and test splits as validation test splits. 

For the ABX task, we also trained the baseline and the regularized CPC models on English, French, and Mandarin datasets. This allows us to evaluate our models across different languages with varying training sizes in a monolingual setting. We also train multilingual systems with data pooled from all three languages together. For the cross-lingual setting, we used the CPC models trained on Librispeech. We evaluated the systems across the three languages for the multilingual and cross-lingual settings. 

\begin{table*}[h!]
    \caption{ABX scores on ZS17 English dataset. CPC models are trained on LibriSpeech 100h portion.}
    \label{tab:abxcpc100}
    \centering
    \begin{tabular}{@{}lcccc|cccc@{}}
    \toprule
    & \multicolumn{4}{c|}{Within}
         & \multicolumn{4}{c}{Across} \\ \cmidrule{2-9}
    & 1s & 10s & 120s & avg & 1s & 10s & 120s & avg \\ \midrule 
    CPC & 7.3 & 7.1 & 7.2 & 7.2 & 10.3 & 9.6 & 9.8 & 9.9 \\
    CPC + SE (ReLU) & 6.6 & 6.5 & 6.8 & 6.6 & 9.4 & 8.9 & 9.2 & 9.2 \\
    CPC + LorR & {5.8} & {5.6} & {6.1} & {5.9} & {8.4} & {8.0} & {8.5} & {8.3} \\
    CPC +  SE (ReLU) + LorR & 6.2 & 5.8 & 6.1 & 6.0 &8.8 & 8.3 & 8.5 & 8.5 \\ 
    CPC-2L + LorR & \textbf{5.6} & \textbf{5.2} & \textbf{5.5} & \textbf{5.5} & \textbf{8.5} & \textbf{7.8} & \textbf{8.1} & \textbf{8.1}\\
    \bottomrule
    \end{tabular}
\end{table*}

\subsection{Architecture Details}
We followed the improved CPC~\cite{riviere2020unsupervised}, which replaces the batch-norm with channel-norm. This helps stabilize the model training and prevents poor solutions. Each of the linear transformation, $W_{m}$, used for predicting $\zvi{t+m}$ from $\cvi{t}$ is now replaced with a single-layer transformer~\cite{vaswani2017attention}. 
This new modified layer can access the entire past till time $t$ to predict $\zvi{t+k}$. Dropout is used in the transformer layers to improve the system's performance. GRU is replaced with LSTM as the recurrent neural network for generating contextual representations. This modification improves the downstream performance of the features extracted from the CPC. 
These modifications allow us to reduce the number of channels in the convolutions layer from 512 to 256 without impacting the performance while significantly reducing the memory requirements. 

In CPC, the encoder was a 5-layer convolutional network with 256 channels in each layer with kernel sizes: 10,8,4,4,4 and strides 5,4,2,2,2. The encoder had a total downsampling factor of 160, with a stride of about 10ms. For a 16kHz input, each feature encodes 10ms of audio and generates 100 features per second. We used a single-layer LSTM with 256 hidden units as the recurrent network. We predict 12 frames into the future, i.e., M=12, and use 128 negative examples. All the models are trained with a batch size of 12 on a single GPU.

\begin{table}[]
    \caption{ABX performance on ZS17 English dataset, The CPC model is pretrained on the Librispeech 360 dataset}
    \label{tab:abxcpc360}
    \centering
    \begin{tabular}{@{}lcc@{}} 
        \toprule 
         & Within & Across \\ \midrule
        \multicolumn{2}{l}{Trained on Zerospeech2017 (45h)} \\ 
        Supervised topline~\cite{dunbar2017zero} & 5.3 & 6.9\\
        Heck et al.~\cite{heck2017feature} & 6.2 & 8.7 \\
        Chorowski et al.~\cite{chorowski2019unsupervised} & 5.5 & 8.0 \\ \midrule 
        \multicolumn{2}{l}{Trained on Librispeech 360} \\
        Original CPC~\cite{oord2018representation} & 9.6 & 13.0 \\
        CPC~\cite{riviere2020unsupervised}  & 6.5 & 8.5\\
        CPC+LorR  & {5.5} & {7.4} \\ 
        CPC-2L+LorR & \textbf{4.7} & \textbf{6.6} \\
        \bottomrule
    \end{tabular}
\end{table}

\begin{table}
    \caption{Linear phone classification accuracy (\%) on Librispeech 100 trained on top of frozen CPC features. CPC models are trained on 100 hours of Librispeech data.}
    \label{tab:prcpc100}
    \centering
    \begin{tabular}{@{}lcc@{}} \toprule
        model & Train Acc & Test Acc \\ \midrule
        CPC & 69.49 & 69.10 \\
        CPC + SE(ReLU) & 70.40 & 69.99 \\
        CPC + LorR & \textbf{71.58} & \textbf{71.18} \\
        CPC + SE(ReLU) + LorR & 71.56 & 71.10 \\ 
        \bottomrule
    \end{tabular}
\end{table}

\begin{table}[]
    \caption{Linear Phone classification accuracy
on the English LibriSpeech-100h dataset. Pretraining data-LS360h}
    \label{tab:prcpc360}
    \centering
    \begin{tabular}{@{}lc@{}}
    \toprule
         &  Test acc \\ \midrule
    Supervised topline~\cite{riviere2020unsupervised} & 76.3 \\ \midrule
    Original CPC~\cite{oord2018representation}  & 65.5 \\ 
    CPC~\cite{riviere2020unsupervised} & 68.9 \\ 
    CPC + LorR & \textbf{71.4} \\
    \bottomrule
    \end{tabular}
\end{table}

\subsection{ABX and Linear Phone Classification}
We evaluated the proposed algorithms on ABX and the phone classification tasks. Table~\ref{tab:abxcpc100} shows the performance of features extracted from baseline CPC and CPC models trained with various regularization techniques on the Librispeech 100 hours portion. All the regularization techniques outperformed the baseline CPC system. Among the different regularization techniques, the LorR worked the best. On average, LorR reduced the ABX error rates by $18\%$ and $16\%$ relative to the baseline in Within and Across conditions, respectively. We also experimented with increasing the number of layers in the autoregressive network, $f_\mathrm{ar}$. CPC-2L denotes the results with two-layer LSTM as $f_\mathrm{ar}$. We observe that with increased model size, the performance improves. 

Table~\ref{tab:abxcpc360} compares the CPC trained on Librispeech 360 dataset with other existing systems on the Zerospeech 2017 English dataset. 
We lowered the ABX scores by $15\%$ and $13\%$ relative to the baseline in Within and Across testing conditions compared to the CPC models. As seen in Table~\ref{tab:abxcpc360},  
our regularized models also outperformed the state-of-art feature extraction methods on the ABX task. One key difference is the amount of data used, we train on a larger amount of data, but it is out-of-domain data. 
We also outperformed the supervised topline, which uses posteriorgrams extracted from the supervised HMM-GMM phone recognition system as features. The models trained on 360 hours of data outperformed the models trained on 100 hours of data.

We also trained a linear phone classifier on top of the representations extracted from the CPC models.  
CPC models were only used as feature extractors and were not finetuned during the phone classifier training. 
We compared the accuracies of various regularization techniques using the CPC models trained on 100 hours of data.
As seen in Table~\ref{tab:prcpc100}, similar to the ABX results, linear phone accuracies for the regularized models outperform the baseline CPC model. The LorR regularization performed best among different regularizations. For the regularized CPC systems, we used the same optimal regularization weights discovered on the ABX task. We compared our best regularized CPC system with the best CPC baseline model in Table~\ref{tab:prcpc360}. We outperformed the CPC model and moved towards matching the topline performance.   

\begin{table}
    \caption{Impact of SE regularization weight $\lambda$ on the ABX scores. CPC model was trained on Librispeech 100 }
    \label{tab:cpcsegrid}
    \centering
    \begin{tabular}{@{}ccc@{}}
    \toprule
    $\lambda$ & Within & Across \\ \cmidrule{2-3}
        0.2 & 6.98 & 9.58 \\
        0.4 & \textbf{6.64} & \textbf{9.17} \\
        0.6 & 6.77 & 9.31  \\
        0.8 & 6.86 & 9.75 \\
        1.0 & 6.94 & 9.64 \\
         \bottomrule
    \end{tabular}
\end{table}

\begin{table}
    \caption{Impact of Window size and LorR loss weight. CPC model was trained on Librispeech 100 }
    \label{tab:cpclorrgrid}
    \centering
    \begin{tabular}{@{}ccc|cc@{}}
    \toprule
    $\alpha$& \multicolumn{2}{c|}{w=2} & \multicolumn{2}{c}{w=3} \\ \cmidrule{2-5}
    & Within & Across & Within & Across\\ \cmidrule{2-5}
         0.2 & 6.07 & 8.65 & 6.35 & 9.20\\
         0.4 & 6.09 & 8.69 & 6.15 & 8.83\\
         0.6 & 5.98 & 8.50 & 6.10 & 8.51\\
         0.8 & 6.09 & 8.57 & \textbf{5.92} & \textbf{8.43}\\
         1.0 & \textbf{5.86} & \textbf{8.29} & 5.94 & 8.71\\ \bottomrule
    \end{tabular}
\end{table}

\begin{table*}
    \caption{ABX scores on ZS17 Mandarin dataset. Training data is shown in parentheses.}
    \label{tab:abxmand}
    \centering
    \begin{tabular}{lcccc|cccc}
    \toprule
    & \multicolumn{4}{c|}{Within}
         & \multicolumn{4}{c}{Across} \\ \cmidrule{2-9}
    & 1s & 10s & 120s & avg & 1s & 10s & 120s & avg \\ \midrule 
    CPC (mand) & 11.3 & 11.2 & 11.3 & 11.3 & 13.6 & 13.3 & 13.6 & 13.5 \\
    CPC + LorR (mand) & 11.4 & 11.1 & 11.1 & 11.2 & 12.0 & 12.2 & 12.3 & 12.2\\
    CPC + LorR (libri2.5) & 12.8 & 12.9 & 13.0 & 12.9 & 15.2 & 15.4 & 15.5 & 15.4 \\
    \rowcolor{Gray}
    CPC (eng+fr+mand) & 11.0 & 10.8 & 11.1 & 11.0 & 11.6 & 11.7 & 12.3 & 11.9 \\
    \rowcolor{Gray}
    CPC + LorR (eng+fr+mand) & 9.9 & 9.6 & 10.1 & 9.9 & 10.1 & 10.0 & 10.8 & 10.3\\
    CPC (libri100) & 10.0 & 9.8 & 9.8 & 9.9 & 9.9 & 9.6 & 9.9 & 9.8\\
    CPC + LorR (libri100) & 8.5 & 8.6 & 9.1 & 8.8 & 8.7 & 8.5 & 9.1 & 8.8\\
    \bottomrule
    \end{tabular}
\end{table*}

\begin{table*}
    \caption{ABX scores on ZS17 French dataset. Training data is shown in parentheses.}
    \label{tab:abxfr}
    \centering
    \begin{tabular}{@{}lcccc|cccc@{}}
    \toprule
    & \multicolumn{4}{c|}{Within}
         & \multicolumn{4}{c}{Across} \\ \cmidrule{2-9}
    & 1s & 10s & 120s & avg & 1s & 10s & 120s & avg \\ \midrule 
    CPC (fr) & 12.5 & 13.1 & 13.2 & 12.9 & 17.8 & 17.5 & 17.9 & 17.7 \\
    CPC + LorR (fr) & 10.5 & 10.7 & 11.0 & 10.7 & 15.5 & 15.0 & 15.5 & 15.3 \\
    CPC + LorR (libri24) & 10.6 & 10.4 & 10.6 & 10.5 & 15.4 & 15.1 & 15.6 & 15.4 \\
    \rowcolor{Gray}
    CPC (eng+fr+mand) & 12.0 & 12.1 & 12.4 & 12.2 & 17.3 & 16.7 & 17.3 & 17.1 \\
    \rowcolor{Gray}
    CPC + LorR (eng+fr+mand) & 10.7 & 10.7 & 11.1 & 10.8 & 15.3 & 14.6 & 15.3 & 15.1\\
    CPC (libri100) & 10.7 & 10.6 & 10.8 & 10.7 & 15.2 & 14.8 & 14.9 & 15.0 \\
    CPC + LorR (libri100) & 9.6 & 9.2 & 10.3 & 9.7 & 13.8 & 13.1 & 13.7 & 13.5\\
    \bottomrule
    \end{tabular}
\end{table*}

\begin{table*}
    \caption{ABX scores on ZS17 English dataset. Training data is shown in parentheses.}
    \label{tab:abxeng}
    \centering
    \begin{tabular}{@{}lcccc|cccc@{}}
    \toprule
    & \multicolumn{4}{c|}{Within}
         & \multicolumn{4}{c}{Across} \\ \cmidrule{2-9}
    & 1s & 10s & 120s & avg & 1s & 10s & 120s & avg \\ \midrule 
    CPC (eng) & 9.2 & 9.4 & 9.6 & 9.4 & 13.1 & 12.9 & 12.9 & 13.0 \\
    CPC + LorR (eng) & 7.5 &  7.5 &  7.8 & 7.6 & 11.0 & 10.8 & 11.0 & 11.0 \\
    CPC + LorR (libri45) & 6.2 & 6.2 & 6.4 & 6.3 & 9.5 & 9.2 & 9.4 & 9.4 \\
    \rowcolor{Gray}
    CPC (eng+fr+mand) & 9.4 & 9.7 & 10.0 & 9.7 & 13.4 & 13.3 & 13.7 & 13.4 \\
    \rowcolor{Gray}
    CPC + LorR (eng+fr+mand) & 7.9 & 8.0 & 8.4 & 8.1 & 11.4 & 11.1 & 11.4 & 11.3\\
    CPC (libri100) & 7.3 & 7.1 & 7.2 & 7.2 & 10.3 & 9.6 & 9.8 & 9.9 \\
    CPC + LorR (libri100)& {5.8} & {5.6} & {6.1} & {5.9} & {8.4} & {8.0} & {8.5} & {8.3} \\
    \bottomrule
    \end{tabular}
\end{table*}

\subsection{Hyperparameter tuning}

For CPC + SE(ReLU) 
experiments, we varied the regularization weight $\lambda$ from 0.2 to 1 with a step size of 0.2. Table~\ref{tab:cpcsegrid} shows the ABX scores with different weights. For the SE(ReLU), the optimal weight was $0.4$.

For the LorR regularization, we have two hyperparameters, the window size $w$ and the regularization weight $\alpha$. We varied the regularization weight $\alpha$ from 0.2 to 1 with a step size of 0.2 and trained CPC systems.  
Table~\ref{tab:cpclorrgrid} shows the ABX scores with different weights. LorR window size two and weight 1.0 performed the best. Please note that all the systems for both SE and LorR, regardless of the choice of weight, outperformed the baseline CPC system.

\subsection{Mono, Cross and Multilingual performance }
We want to evaluate the performance of the proposed regularization in mono, cross, and multilingual setting. We used Zerospeech 2017 dataset, which contains three languages: English, French, and Mandarin. These three include 45, 24, and 2.5 hours of data, respectively. 

We trained the baseline CPC and CPC for each language with LorR regularization. As seen in Tables \ref{tab:abxmand}, \ref{tab:abxfr} and \ref{tab:abxeng}, we see consistent improvements in both within-speaker and across-speaker conditions across all the languages on average.  
For Mandarin (Table \ref{tab:abxmand}), we see the slightest improvement in the within-speaker condition. In 1s evaluation, the CPC baseline performed slightly better (11.3) than the CPC + LorR system (11.4). The scores before rounding are 11.34 and 11.36, respectively. In the across-speaker condition, we observe consistent improvements similar to other languages. 

For the multilingual setting, we pooled the data from all three languages and trained the baseline CPC and CPC + LorR system on 71.5 (45 + 24 + 2.5) hours of data. The CPC + LorR outperformed the baseline CPC in all three languages in the multilingual setting. 
We observed that the Multilingual system outperformed the monolingual system for Mandarin (Table \ref{tab:abxmand}). We observed little to no performance improvement for French (Table \ref{tab:abxfr}). However, for English (Table \ref{tab:abxeng}), the multilingual system performed worse than the monolingual system. We think this is because Mandarin has the smallest amount of data. French and English have a sufficiently large amount of data to train a decent monolingual system. 

For the cross-lingual setting, we used the CPC and CPC + LorR system trained on 100 hours portion from the Librispeech dataset. There is a mismatch in the training language and dataset for Mandarin and French, i.e., training is done on Librispeech, whereas testing is done on the Zerospeech 2017 dataset. For English, the only mismatch is in the datasets. As observed in Tables \ref{tab:abxmand}, \ref{tab:abxfr}, and \ref{tab:abxeng}, the CPC + LorR system outperforms the CPC system across all languages. 
The CPC/ CPC + LorR trained on Librispeech 100 hours of data outperformed the system trained on individual languages or in a multilingual setting (Tables \ref{tab:abxmand}, \ref{tab:abxfr} and \ref{tab:abxeng}). We believe this is because of the amount of data. The total data is less than 100 hours, even in a multilingual setting.

To test our hypothesis, we train three systems on 2.5, 24, and 45 hours subsets from Librispeech 100 hours dataset. 
For Mandarin (Table \ref{tab:abxmand}), the CPC + LorR trained on monolingual data, i.e., Mandarin, outperforms the cross-lingual system trained on Librispeech with the same amounts of data. For French (Table \ref{tab:abxfr}), the monolingual and cross-lingual CPC + LorR systems perform similarly. Increasing the amount of Librispeech data for both languages improves the performance. For English(Table \ref{tab:abxeng}), we observe the CPC + LorR system trained on Librispeech 45 hours subset outperforms the system trained on Zerospeech English data (45 hours). Increasing the amount of Librispeech data further improves the performance. 

These experiments show the language invariance of the proposed regularization. Across all the training conditions (monolingual, cross-lingual, and multilingual) and various languages (Mandarin, French, and English), CPC + LorR constantly outperformed the baseline CPC. 

\subsection{Are SE and LorR Complementary?}

We want to analyze whether the regularization techniques, i.e., SE and LorR, are complementary and whether we can use them simultaneously to improve performance further. We train a CPC model which uses both SE(ReLU) and LorR regularization techniques. The total loss of the model is given as
\begin{equation}
    \mathcal{L}_{\textrm{total}} = \mathcal{L}_{\textrm{CPC}} + 0.5(\mathcal{L}_{\textrm{LorR}} + 0.4\mathcal{L}_{\textrm{SE(ReLU)}} ) \;.
\end{equation}
The weights of $\mathcal{L}_{\textrm{LorR}}$ and $\mathcal{L}_{\textrm{SE(ReLU)}}$ are the best weights from the individual systems. 

We evaluated this system on ABX and linear phone classification. As seen in Tables~\ref{tab:abxcpc100} and \ref{tab:prcpc100}, it performs better than CPC + SE system but worse than CPC + LorR system. Either the regularization techniques do not have complementary information, or the regularization weights used might not be the optimal choice.  

\begin{table}[h!]
\caption{Linear phone classification results on Librispeech 100 trained on top of concatenated features from frozen CPC models. CPC models are trained on 100 hours of Librispeech data.}
\label{tab:cpc100concat}
\centering
\begin{tabular}{@{}ccccc@{}}
\hline
CPC & CPC + SE & CPC + LorR & Train Acc & Test Acc \\ \hline
\cmark & - & - & 69.49 & 69.10 \\
\cmark & \cmark & - & 72.35 & 71.82 \\
\cmark & - & \cmark & 73.04 & 72.54 \\
- & \cmark & \cmark & 73.39 & 72.84 \\
\cmark & \cmark & \cmark & 74.08 & 73.54 \\ \hline
\end{tabular}
\end{table}

To find the optimal choice for weighing the SE and LorR regularization, we need to train multiple systems. However, that is computationally expensive. We, therefore, simply concatenate the features extracted from different models and train a linear phone classifier.
Table~\ref{tab:cpc100concat} shows phone classification accuracy on various system combinations. The phone classifier trained on top of concatenated features from CPC + SE(plus1) and CPC + LorR performed the best among the two system combinations. This suggests that complementary information is present in the two features, improving the performance.

\subsection{Clustering Analysis}

Unsupervised features are often used for acoustic unit discovery (AUD). In AUD, speech signals are segmented and clustered into discrete phone-like units. We want to test how the representation produced by the regularized CPC cluster compared to baseline CPC. We ran K-means on top of the latent representation extracted from the frozen CPC models. We trained the K-means algorithm on the 100 hours of Librispeech data. We experimented with varying numbers of clusters since the actual number of clusters is unknown. 

We use purity and Normalized mutual information (NMI), two commonly used metrics for measuring the clustering quality. As seen from Table~\ref{tab:cpc100clust}, the results for CPC + LorR consistently outperformed the baseline CPC. This means the clusters generated by CPC + LorR aligned better with the forced-aligned phone labels than the baseline CPC. 

\begin{table}[h!]
    \caption{Purity and NMI (in parenthesis) on the Librispeech test split for a different number of clusters.  }
    \label{tab:cpc100clust}
    \centering
    \begin{tabular}{@{}cccc@{}} \toprule
         & 25 & 50 & 100 \\ \cmidrule{2-4}
        CPC & 40.5 (38.5) & 43.6 (37.7) & 46.8 (37.2) \\ 
        CPC+LorR & \textbf{41.2 (40.0)} & \textbf{45.2 (40.2)} & \textbf{51.2 (40.6)} \\
        \bottomrule
    \end{tabular}
\end{table}

\subsection{Original CPC and regularization}

In the original CPC, a linear transformation $W_{m}$ is used for predicting $\zvi{t+m}$ from $\cvi{t}$. In the modified implementation, a single-layer transformer is used for predicting $\zvi{t+m}$. 
Each transformer layer introduces around 1.3 million new parameters, and there are twelve one-layer transformers in total. The new CPC variant improves the performance significantly but also adds more parameters to the model. We want to see if our regularization works with the old CPC variant.

The regression weights are not used during inference but only during training. The inference model is the same size for the original and the modified variant. As observed in Table~\ref{tab:origCPCregu}, our regularized CPC outperformed the original variant. However, the performance was still worse than the modified version of CPC, the original version contained around 10$\%$ parameters, which shows the usefulness of the extra parameters. 

\begin{table}[h!]
    \caption{Performance comparison for the original and the modified implementation of the CPC}
    \label{tab:origCPCregu}
    \centering
    \begin{tabular}{@{}lccc@{}}
        \toprule
        & parameters & Within & Across \\ 
        \cmidrule{2-4}
        Original CPC & 1.8m & 8.6 & 12.2\\
        Original CPC + LorR & 1.8m & 7.6 & 11.0\\
        CPC & 17.6m &  7.2 & 9.9\\
        CPC + LorR & 17.6m & 5.9 & 8.3 \\ \bottomrule
    \end{tabular}
\end{table}

\subsection{Probing for Speaker Information}

As seen from the ABX results~\ref{tab:abxcpc100}, the performance across-speaker testing conditions are much worse than within-speaker conditions. This implies the features still contain speaker information. Here, we want to analyze how much speaker information is present in the representations. We followed~\cite{van2021analyzing} and trained a linear classifier that tries to predict the identity of the speaker from a single frame. We used the last frame from the GRU as the input. It summarizes all the information in the sequence. The speaker classifier was trained for 50 epochs and evaluated on the same train test split used for the linear phone classifier on Librispeech 100.

As seen from Table~\ref{tab:spkcls}, the speaker classification performance for CPC + LorR is much lower than the baseline CPC. That implies the features extracted from CPC + LorR suppressed speaker information. Even on Mandarin monolingual training~\ref{tab:abxmand} where we get a small improvement (0.1) in the within speaker condition, in the across speaker testing condition, we see good improvement (1.3). We believe this is due to regularization removing some speaker information from the features.  

\begin{table}[]
    \caption{Speaker Classification performance on Librispeech 100}
    \label{tab:spkcls}
    \centering
    \begin{tabular}{@{}ll@{}}
    \toprule
        & Acc \\ \cmidrule{2-2}
        CPC  &  77.0 \\ 
        CPC + LorR & 38.8 \\ \bottomrule
    \end{tabular}
\end{table}

\subsection{Data Augmentation for CPC}
Data augmentation techniques greatly improve the performance of the model in supervised scenarios, especially in limited labeled data scenarios. Data augmentation has also become a significant part of SSL systems. Typically, a signal is augmented to generate two views, and the feature extractor is trained to generate the same representation regardless of the augmentation. 

The idea is that the underlying class information remains the same regardless of the augmentation. The goal is to learn augmentation invariant representation. By doing so, the representations would capture the underlying class information. 

CPC + Aug~\cite{kharitonov2021data} proposed augmentation of the speech signal in the time domain to improve the CPC performance. At a given point $t$, the past audio signal, i.e., audio from the beginning till the time $t$, is fed into the convolutional encoder and then into the autoregressive network to generate the context $\cvi{t}$. The context $\cvi{t}$ is then used to predict future feature frames $\zvi{t+k}$. 

\begin{figure}
    \centering
    \includegraphics[width=2.5in]{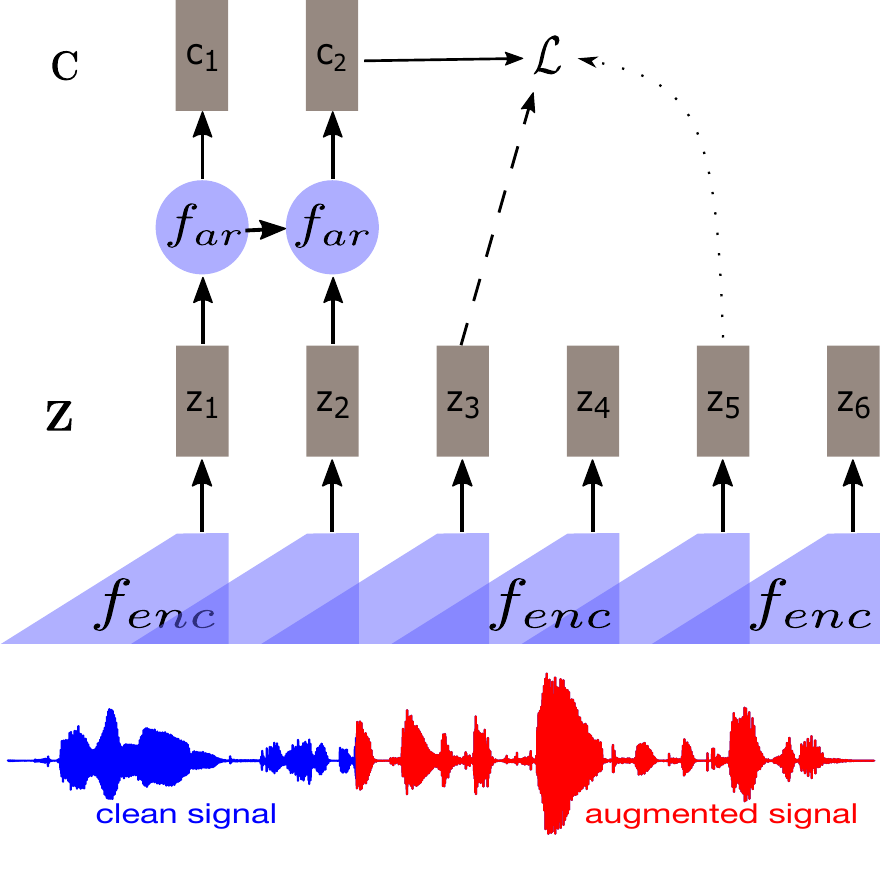}
    \caption{Overview of the CPC architecture. The solid line represents the reference frame; the dashed line shows the positive, and the dotted line shows the randomly sampled negative example.}
    \label{fig:augcpc}
\end{figure}

In the augmented CPC, the past audio signal is corrupted with a noise used for generating the context. This context is then used to predict the feature frames generated by future audio signals corrupted with a different augmentation. This forces the encoder to denoise the data and generate a similar representation regardless of the augmentation. They showed consistent improvements over the baseline CPC. They experimented with augmenting both past and future while training the CPC. Experimentally only augmenting the future while using the unaugmented speech data for the past performed best. The final architecture is shown in Figure~\ref{fig:augcpc}.

CPC + Aug experimented extensively with different types of augmentation, their relative order of application, and which portion of the audio signal should be augmented. We followed the experiments in ~\cite{kharitonov2021data} and used pitch shift, noise addition, and reverberation to augment the audio signal. For pitch: we changed the pitch by an integer uniformly sampled between -300 and 300(the change value is measured by 1/100 of a tone). For the additive noise, we used the MUSAN dataset. The additive noise was filtered through a bandpass filter in the [80, 240] Hz range. For reverb, the room scale was randomly sampled between 0 and 100, and all other parameters were fixed to defaults. We only augment the future audio signal, i.e., both positive and negative samples were generated from the augmented signal. We used WavAugment provided by ~\cite{kharitonov2021data} for the data augmentation experiments. 

They also modified the architecture of the CPC: the Ws, instead of being modeled by one layer transformer independently, were all now modeled with a single transformer with $M(=12)$ classification heads. This improved the training time without significant reductions in performance. For the recurrent context network, a two-layer LSTM is used. For the data augmentation experiments, we follow the same architecture for a fair comparison. 

To avoid reading the MUSAN noises in every batch, we preloaded random three seconds noise samples from each noise utterance in MUSAN. In an ideal case, we should load all the utterances from the MUSAN dataset, but that is very memory intensive. 
While augmenting the speech signal, we randomly sampled a 1.28 seconds signal from the three seconds noise samples and added it to the speech signal. We reload the three seconds noise samples from MUSAN every 100k update steps. 

One important question for the augmentation is how much we should augment. We carried out experiments to discover if it is a good idea to augment all the time or augment some times and use unaugmented data the rest of the time. We started by using only unaugmented and increased the probability of using augmented data in steps of 0.2 to using only augmented data. As observed in Table~\ref{tab:augprob}, augmenting all the time can be detrimental to the system's performance. 
We hypothesize that using clean data allows the model to focus on just feature extraction instead of feature extraction and denoise. Using augmented data all the time still outperforms just using clean data.

The detailed experimental results of our proposed regularization method and the augmented CPC baseline are in Tables~\ref{tab:abxaugall}. As evident from the table, regularization and augmentation can be complementary, and we see consistent improvements over the augmented CPC baseline across all three languages. 

Our implementation of the augmented CPC performed better than the reported number in the augmented CPC paper. We believe this is due to the following differences: our strategy for additive noise is different. We effectively sample from a smaller pool of additive noises. 
There might also be differences in the number of training epochs for the reported numbers for the augmented CPC system vs. our implementation. 

\begin{table}[]
    \caption{probability of using unaugmented data. CPC trained on ZS17 Mandarin (2.5h) and tested on Mandarin test set }
    \label{tab:augprob}
    \centering
    \begin{tabular}{@{}ccc@{}}
        \toprule
        & within & across \\ \cmidrule{2-3}
        0.0 & 10.2 & 9.2 \\
        0.2 & 9.9 & 8.8 \\
        0.4 & \textbf{9.5} & \textbf{8.8} \\
        0.6 & 9.8 & 9.1 \\
        0.8 & 10.6 & 10.6 \\ 
        1.0 & 11.2 & 12.2 \\ \bottomrule
    \end{tabular}
\end{table}

\begin{table}[]
    \caption{ABX scores on Zerospeech 2017 challenge. The models are trained on 45, 24, and 2.5 hours of English, French, and Mandarin data, respectively. ``W" and ``A" denotes the within and across speaker scores.}
    \label{tab:abxaugall}
    \centering
    \begin{tabular}{lcccccc}
        \toprule
    & \multicolumn{2}{c}{English}
         & \multicolumn{2}{c}{French} 
         & \multicolumn{2}{c}{Mandarin} \\ 
         \cmidrule{2-7}
         & W & A & W & A & W & A \\ \cmidrule{2-7}
         CPC (2L) + Aug~\cite{kharitonov2021data} & 6.6 & 9.3 & 9.3 & 14.1 & 11.2 & 11.9 \\
         CPC (2L) + Aug & 6.4 & 8.8 & 9.1 & 13.1 & 9.8 & 9.6 \\
         CPC (2L) + Aug + LorR & \textbf{6.0} & \textbf{8.5} & \textbf{8.8} & \textbf{12.6} & \textbf{9.5} & \textbf{8.8} \\ 
         \bottomrule
    \end{tabular}

\end{table}

\subsection{Zerospeech 2021 challenge}

The 2021 version of the Zerospeech challenge~\cite{dunbar2021zero} used Librispeech for training and evaluation for the ABX task. We compare our system with the small budget CPC baseline, which was trained on Librispeech 100 hours subset. Another method we compared is ACPC~\cite{chorowski2021aligned}, which improves over the CPC system. All the systems use two-layer LSTMs in the $f_{\mathrm{ar}}$.

Multilevel systems such as mACPC~\cite{cuervo2022contrastive} and HCPC~\cite{cuervo2022variable} outperform CPC and ACPC but also introduce more parameters and complexity to the model. For example, the architecture for the lower CPC in HCPC is the same as our models, but it also has additional layers for the higher level CPC, the boundary detector, etc. It also requires finetuning the reinforcement learning-based boundary detector and the quantization module. 

Our method is more straightforward and does not introduce more parameters to CPC. As seen in Table~\ref{tab:abxzs21}, our systems outperforms the existing single-level systems, such as CPC~\cite{wang2019transformer} and ACPC~\cite{chorowski2021aligned}, and multilevel systems, such as mACPC~\cite{cuervo2022contrastive} and HCPC~\cite{cuervo2022variable}. 
Adding augmentation further improves the system performance, especially in the across-speaker setting. 

\begin{table}[]
    \caption{ABX performance on ZeroSpeech 21 English dev set.}
    \label{tab:abxzs21}
    \centering
    \begin{tabular}{@{}lcccc@{}} 
        \toprule 
         &  \multicolumn{2}{c}{Within} &  \multicolumn{2}{c}{Across} \\ \midrule
         & clean & other & clean & other \\ \midrule
        CPC (2L) & 6.18 & 8.46 & 8.02 & 13.59 \\
        ACPC (2L) & 5.37 & 7.46 & 7.09 & 12.60 \\
        mACPC (2L) & 5.13 & -& 6.84 & -\\
        HCPC (2L) & 5.08 & -& 6.72 & -\\
        CPC + LorR (2L) & 5.05 & 7.16 & 6.48 & 11.90 \\ 
        CPC + LorR + Aug (2L) & \textbf{4.90} & \textbf{6.88} & \textbf{6.17} & \textbf{10.96} \\
        \bottomrule
    \end{tabular}
\end{table}

\subsection{Cross lingual Phone recognition}

CPC models are commonly used as feature extractors for downstream tasks such as automatic speech recognition. As evident from the linear phone classification experiments, CPC features capture the phoneme information quite well. We want to analyze how well these features work in different languages. We follow the experiments in ~\cite{riviere2020unsupervised} and consider the task of phoneme classification on different languages from the common voice dataset. Phoneme recognizers are trained with CTC objective.

To show the advantage of using pretrained feature extractors in a low-resource setting, we used the model trained only on the 1-hour target dataset. Table \ref{tab:crossASR} shows the phone error rates on the CommonVoice dataset. The models trained from scratch perform poorly. The models trained on top of the pretrained feature extractor work significantly better across all languages. In both 100 and 360 hours training settings, our regularized CPC models outperformed the baseline CPC models. Our regularized CPC systems trained with just 100 hours of unlabeled data almost matched the performance of CPC trained with 360 hours of data. 

Unsupervised feature extractors typically require more data to match the performance of supervised pre-trained feature extractors~\cite{schneider2019wav2vec,riviere2020unsupervised}. While CPC needed 360 hours of unlabeled data to match the quality of supervised pretraining, our regularized CPC matched the performance with just 100 hours of data. With 360 hours of data, our models outperformed the supervised pretraining, which is significant as it is easier to collect unlabeled data than labeled data. 

\begin{table*}[h!]
\caption{Cross lingual transfer of pretrained features in terms of phone error rates. Training: 1h data/language from Common voice. The languages are: Dutch (du), Spanish (es), French (fr), Italian (it), Kyrgyz (ky), Russian (ru), Sweedish (sv), Turkish (tr), Tatar (tt) and Mandarin (zh)}
\label{tab:crossASR}
\centering
\begin{tabular}{@{}lccccccccccccc@{}}
\toprule
Model &Pretraining &Frozen &du &es &fr &it &ky &ru &sv &tr &tt &zh &Avg \\ \midrule
From scratch &- &No &84.7 &95.9 &95.1 &95.0 &81.5 &97.7 &86.1 &83.1 &72.9 &84.3 &87.6 \\
Bottleneck &Babel-1070h &Yes &47.9 &36.6 &48.3 &39.0 &38.7 &45.2 &52.6 &43.4 &42.5 &54.3  &44.9 \\
Supervised &LS-100h &Yes &42.4 &36.4 &47.0 &40.5 &41.0 &43.6 &47.0 &48.5 &41.5 &56.8 &44.5 \\ \midrule
Orignal CPC &LS-100h &Yes &51.5 &44.2 &54.5 &47.0 &44.8 &49.0 &54.0 &54.7 &48.9 &60.1 &50.9 \\
CPC &LS-100h &Yes &44.4 &38.7 &49.3 &42.1 &40.7 &45.2 &48.8 &49.7 &44.0 &55.5 &45.8 \\
CPC + LorR & LS-100h & Yes & 45.7 & 36.5 & 45.9 & 42.1 & 40.1 & 46.5 & 48.9 & 48.0 & 37.4 & 55.3 & 44.6 \\ 
CPC & LS-360h &Yes &\textbf{42.5} &38.0 &47.1 &\textbf{40.5} &41.2 &\textbf{43.7} &\textbf{47.5} &47.3 &42.0 &55.0 &44.5 \\
CPC + LorR & LS-360h & Yes & 43.3 & \textbf{35.0} & \textbf{43.6} & 40.7 & \textbf{38.9} & 44.5 & 47.7 & \textbf{46.1} & \textbf{35.6} & \textbf{54.0} & \textbf{42.9}\\
\bottomrule
\end{tabular} 

\end{table*}

\section{Conclusions and Future work}
SSL methods such as CPC have become the front end of speech technologies. ASR systems trained on top of SSL features require much less labeled data to achieve the same performance as models trained from speech data. We can further lower the labeled data requirements by improving the feature extraction methods. In this work, we propose regularization techniques that impose slowness constraints on the features learned by a CPC model. We compared our proposed methods with the CPC baseline on ABX, linear phone classification,  clustering, and phoneme recognition tasks. Our models outperformed the baseline CPC models on all the tasks in monolingual, cross-lingual, and multilingual settings. Left-or-Right regularization performs the best among the proposed regularizations. 

In the future, we would like to apply these techniques on a larger scale, i.e., training on all of Librispeech. We would also like to use the proposed regularization techniques for other feature extraction techniques, such as Wav2Vec2.0. Multistage methods such as SCPC and mACPC apply segmental constraints via a second-level CPC. We would also like to see if our regularization techniques can improve the multilevel systems discrimination performance. Another interesting direction would be to apply the proposed constraints in a semi-supervised or supervised scenario.

\bibliographystyle{ieeetr}
\bibliography{ref} 

\end{document}